\documentstyle[twoside,fleqn,npb,epsfig]{article}
\newcommand{\AmS}{{\protect\the\textfont2
  A\kern-.1667em\lower.5ex\hbox{M}\kern-.125emS}}

\title{Virtual Photon Structure Functions
\thanks{Presented at RADCOR/Loops and Legs 2002, Kloster Banz, Germany,
September 8-13, 2002. KUCP-222.}}
\author{T. Uematsu\address{Dept. of Fundamental Sciences,
        FIHS, Kyoto University, 
        Kyoto 606-8501, Japan}%
        }

\begin{document}

\begin{abstract}
We discuss the perturbatively calculable virtual photon structure
functions. First we present the framework for analyzing the structure
functions of the virtual photon and derive a first moment 
of $g_1^\gamma$ of the virtual photon.
We then investigate the three positivity constraints satisfied by
the eight structure functions of the virtual photon.

\end{abstract}

\maketitle

\section{INTRODUCTION}

I would like to talk about the virtual photon structures,
especially about the general virtual photon structure functions and their
positivity constraints. This work has been done in collaboration
with Ken Sasaki and Jacques Soffer \cite{SSU02}.

The structure of the virtual photon can be studied from
the double-tagged two-photon processes in  
$e^+$$e^-$ collisions (Fig.1) or the 2-jet events in resolved
photon processes of  $e^{-(+)}p$ collisions.

\vspace{-1cm}

\begin{figure}[htbp]
\begin{center}
\epsfxsize=5cm
\hspace*{0cm}
\ \epsfbox{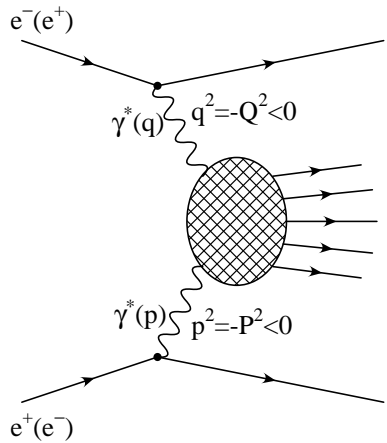}
\vspace{-1cm}
\caption{Two-photon process in $e^+e^-$ collision}
\end{center}
\end{figure}
\vspace*{-1cm}

The experimental data provide the effective photon structure functions, 
 $F_{\rm eff}^\gamma\simeq F_{TT}^\gamma
+F_{TL}^\gamma+F_{LT}^\gamma+F_{LL}^\gamma
\simeq \frac{1}{x}F_2^\gamma+\frac{3}{2}F_L^\gamma$, 
where the indices $T$ and $L$ refer to the transverse and longitudinal
photon, respectively. So far mostly studied are 
the unpolarized photon structure functions,
$F_2^\gamma$ and $F_L^\gamma$.
Here I refer to the recent review articles on virtual photon
structure functions, M. Krawczyk's talk at PHOTON 2000 \cite{Kraw00}, 
and M. Nisius's summary talk at PHOTON 2001 \cite{Nisi01}. More recent and
detailed reviews are those by M. Klasen \cite{Klas02} and by I. Schienbein
\cite{Schi02}.

Now in the last few years, there has been much theoretical
interest in the photon's spin structure functions
$g_1^\gamma$ and $g_2^\gamma$. Especially, $g_1^\gamma$
has attracted much attention because of its relevance
to the axial anomaly just like the case of 
nucleon spin structure function.
Namely we can write down the QCD prediction of the 1st
moment of $g_1^\gamma$. These spin dependent photon structure
functions can be studied from the polarized 
$ep$ collider or more directly from the polarized $e^+e^-$ collision.

The virtual photon
target provides a good place to study the structure of the
operators appearing in the operator product 
expansion of the current product, since the
virtual photon matrix elements of the operators are 
perturbatively calculable.

\section{STRUCTURE FUNCTIONS}

Now there are 8 independent structure functions
for the virtual photon-photon scattering illustrated
in Fig.2. This process is described by the structure tensor


\begin{figure}[htbp]
\begin{center}
\epsfxsize=5cm
\hspace*{0cm}
\ \epsfbox{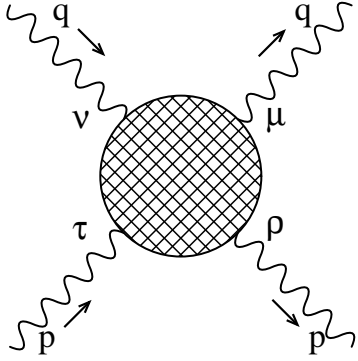}
\vspace{-1cm}
\caption{Virtual photon-photon scattering}
\end{center}
\end{figure}
\begin{equation}
\qquad W_{\mu\nu\rho\tau}(p,q)=\frac{1}{\pi}{\rm Im}\ 
T_{\mu\nu\rho\tau}(p,q)
\end{equation}
where $T_{\mu\nu\rho\tau}$ is the amplitude for
$\gamma(q)+\gamma(p) \rightarrow \gamma(q)+\gamma(p)$
given by
\begin{eqnarray}
&&T_{\mu\nu\rho\tau}(p,q)=i\int d^4xd^4yd^4ze^{iq\cdot x}
e^{ip\cdot(y-z)}\nonumber\\
&& \qquad \times
\langle 0|T\left[J_\rho(y)
J_\mu(x)J_\nu(0)J_\tau(z)\right]|0\rangle
\end{eqnarray}
where $J$ is the electromagnetic current, and $q$ and $p$ are
four-momenta of the probe and target photon, respectively.

For the kinematical region $\Lambda^2 \ll P^2 \ll Q^2$
($\Lambda$ : QCD scale parameter)
or
$x^2 \ll (y-z)^2$ we have the Operator Product Expansion (OPE):
\begin{equation}
J_\mu(x)J_\nu(0)\sim \sum_n C_n(x)O_n(0)
\end{equation}
where $O_n(0)$ and $C_n(x)$ are the spin-$n$ operator
and its coefficient function, respectively.

Now we can decompose the structure tensor following
Budnev, Chernyak and Ginzburg \cite{BCG71} as
\begin{eqnarray}
&&\hspace{-0.7cm}W^{\mu\nu\rho\tau}(p,q)=(P_{TT})^{\mu\nu\rho\tau}W_{TT}+(P_{TT}^a)^{\mu\nu\rho\tau}
W_{TT}^a\nonumber\\
&&\hspace{1.3cm}+(P_{TT}^\tau)^{\mu\nu\rho\tau}W_{TT}^\tau+
(P_{ST})^{\mu\nu\rho\tau}W_{ST}\nonumber\\
&&\hspace{1.3cm}+(P_{TS})^{\mu\nu\rho\tau}W_{TS}+(P_{SS})^{\mu\nu\rho\tau}W_{SS}\nonumber\\
&&\hspace{1.3cm}-(P_{TS}^\tau)^{\mu\nu\rho\tau}W_{TS}^\tau-
(P_{TS}^{\tau a})^{\mu\nu\rho\tau}W_{TS}^{\tau a}\nonumber\\
\label{structurefn}
\end{eqnarray}
where $P_{i}$'s are the projectors given by  \cite{SSU02,BCG71}
\begin{equation}
(P_{TT})^{\mu\nu\rho\tau}=R^{\mu\nu}R^{\rho\tau} \quad \mbox{etc.}
\end{equation}
with
\begin{eqnarray}
&&R^{\mu\nu}=-g^{\mu\nu}+\frac{1}{X}\left[
w(q^\mu p^\nu+p^\mu q^\nu)\right.\nonumber\\
&&\hspace{2.5cm}\left.-q^2 p^\mu p^\nu\ -p^2 q^\mu q^\nu\right]\nonumber\\
&&w=p\cdot q, \quad X=(p\cdot q)^2-p^2q^2
\end{eqnarray}
and satisfy
\begin{eqnarray}
&&(P_{TT})\cdot (P_{TT})=4 \quad \mbox{etc.}\nonumber\\
&&(P_i)^{\mu\nu\rho\tau}(P_j)_{\mu\nu\rho\tau}=0\quad
(i\neq j)
\end{eqnarray}
where the subscripts $T$ and $S$ refer to the transverse
and longitudinal photon, respectively.

We now introduce the s-channel helicity amplitudes 
as follows:
\begin{equation}
W(ab|a'b')=\epsilon^*_{\mu}(a)\epsilon^*_{\rho}(b)
W^{\mu\nu\rho\tau}\epsilon_{\nu}(a')\epsilon_{\tau}(b')
\end{equation}
where $\epsilon_{\mu}(a)$ represents the photon polarization 
vector with helicity $a$, and $a=0,\pm 1$. From angular momentum 
conservation, parity conservation, and time reversal invariance 
\cite{BLS80}, we have in total
8 independent helicity amplitudes, which we may take as
\begin{eqnarray}
&&\hspace{-0.7cm}
W(1,1|1,1), \ W(1,-1|1,-1), \ W(1,0|1,0),\nonumber\\
&&\hspace{-0.7cm}
W(0,1|0,1), \ W(0,0|0,0), \nonumber\\
&&\hspace{-0.7cm}
W(1,1|-1,-1), \ W(1,1|0,0), \ W(1,0|0,-1).
\label{helicityamp}
\end{eqnarray}
The first 5 amplitudes are helicity-nonflip, while the
last 3 amplitudes are helicity-flip. 

Here we note the relation between the 8 structure functions
(\ref{structurefn}) 
and the 8 helicity amplitudes (\ref{helicityamp}), 
which are given by
\begin{eqnarray}
&& W_{TT}=\frac{1}{2}\left[ W(1,1|1,1)+W(1,-1|1,-1) \right]\nonumber\\
&& W_{ST}=W(0,1|0,1)\nonumber\\
&& W_{TS}=W(1,0|1,0), \quad W_{SS}=W(0,0|0,0) \nonumber\\
&& W_{TT}^a=\frac{1}{2}\left[ W(1,1|1,1)-W(1,-1|1,-1) \right]\nonumber\\
&& W_{TT}^\tau=W(1,1|-1,-1)\nonumber\\
&& W_{TS}^\tau=\frac{1}{2}\left[ W(1,1|0,0)-W(1,0|0,-1) \right]\nonumber\\
&& W_{TS}^{\tau a}=\frac{1}{2}\left[ W(1,1|0,0)+W(1,0|0,-1) \right].
\end{eqnarray}
Since the helicity-nonflip amplitudes are non-negative, 
the first four structure functions are postive-definite, while
the last four are not. 
We also note that all structure functions are
symmetric under intechange of $p \leftrightarrow q$ except
$W_{ST}$ and $W_{TS}$, which satisfy
\begin{equation}
W_{ST}(w,q^2,p^2)=W_{TS}(w,p^2,q^2).
\end{equation}

\section{$F_2^\gamma(x,Q^2,P^2)$ and $g_1^\gamma(x,Q^2,P^2)$}

We now turn to 
the NLO QCD
calculation of the virtual photon structure functions
$F_2^\gamma(x,Q^2,P^2)$ and $g_1^\gamma(x,Q^2,P^2)$.
The basic theoretical framework for the NLO QCD calculation
is either based on the OPE supplemented by Renormalization 
Group (RG) or on the DGLAP type parton evolution equation. 
The master formula for the $n$-th moment for $F_2^\gamma$ and $g_1^\gamma$
has the same form, and for the $g_1^\gamma$ case, we have  
 for the kinematical region $\Lambda^2 \ll P^2 \ll Q^2$: 
\begin{eqnarray}
&&\hspace{-0.5cm}\int_0^1 dx x^{n-1}g_1^\gamma(x,Q^2,P^2)=  
\frac{\alpha}{4\pi}\frac{1}{2\beta_0} \times \nonumber\\
&&\hspace{-0.5cm}\Bigl[
\sum_{i=+,-,NS}
L_i^n
\frac{4\pi}{\alpha_s(Q^2)}
\Bigl\{1-\left(\frac{\alpha_s(Q^2)}{\alpha_s(P^2)}\right)^{\lambda_i^n/2\beta_0
+1}\Bigr\}\nonumber\\
&&+\hspace{0.5cm}\sum_{i=+,-,NS}{\cal A}_i^n\Bigl\{1-\left(\frac{\alpha_s(Q^2)}
{\alpha_s(P^2)}\right)^{\lambda_i^n/2\beta_0}\Bigr\}\nonumber\\
&&+\hspace{0.5cm}\sum_{i=+,-,NS}{\cal B}_i^n\Bigl\{1-\left(\frac{\alpha_s(Q^2)}
{\alpha_s(P^2)}\right)^{\lambda_i^n/2\beta_0+1}\Bigr\}\nonumber\\
&&\vspace{2cm}\hspace{1.3cm}+\qquad {\cal C}^n +\quad {\cal O}(\alpha_s) \qquad
 \Bigr]
\label{master}
\end{eqnarray}
where $L_i^n$, ${\cal A}_i^n$, ${\cal B}_i^n$ and ${\cal C}^n$
are the renormalization scheme independent coefficients computed 
from the one- and two-loop anomalous dimensions together with one-loop
coefficient functions.

\vspace*{-4cm}
\begin{figure}[htbp]
\vspace{-0.3cm}
\begin{center}
\epsfxsize=8cm
\hspace*{-0.7cm}
\ \epsfbox{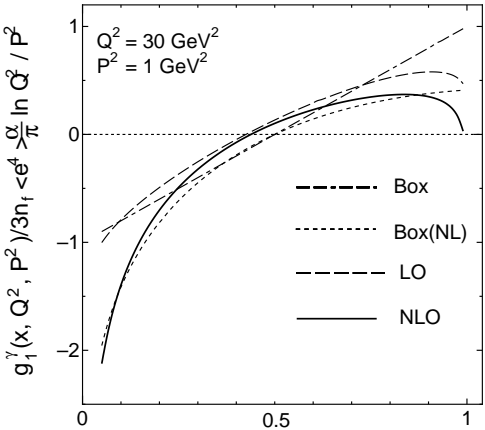}
\vspace{-3.5cm}
\caption{Spin structure function $g_1^\gamma(x,Q^2,P^2)$
for $Q^2=30$ GeV$^2$ and $P^2=1$ GeV$^2$}
\end{center}
\end{figure}
\vspace{0.1cm}
\noindent
$\alpha_s(Q^2)$ is the QCD running coupling constant, and 
$\lambda_i^n \ (i=+,-,NS)$
denote the eigenvalues of one-loop anomalous dimension matrix.
We have shown $g_1^\gamma(x,Q^2,P^2)$ 
in Fig.3, where the NLO $g_1^\gamma$ corresponds to the solid
line, for $n_f=3$, $Q^2=30$ GeV$^2$ and $P^2=1$ GeV$^2$ with 
$\Lambda=0.2$ GeV.
 
For a real photon ($P^2=0$),  
the 1st moment vanishes to all orders of $\alpha_s(Q^2)$ in QCD
\cite{BBS98}:
\begin{equation}
\int_0^1 dx g_1^\gamma(x,Q^2)=0. \label{real}
\end{equation}
Now the question is what about the $n=1$ moment of the virtual photon case.
Taking $n\rightarrow 1$ limit of (\ref{master})
the first 3 terms vanish as
\begin{equation}
L_i^n \rightarrow 0, \quad
\sum_i {\cal A}_i^n\{ \ \} \rightarrow 0, \quad
\sum_i {\cal B}_i^n\{ \ \} \rightarrow 0.
\end{equation}
Denoting $\langle e^4 \rangle=\sum_{i=1}^{n_f} e_i^4/n_f$ 
($e_i$: the $i$-th quark charge, $n_f$: the number of active flavors), we have
\begin{equation}
{\cal C}^{n=1}=12\beta_0\langle e^4 \rangle (B_G^n+A_{qG}^n)|_{n=1}.
\end{equation}
Here we note that the sum of the one-loop coefficient function $B_G^n$
and the finite photon matrix element of quark operator $A_{qG}^n$ is 
renormalization-scheme independent and equal to $-2n_f$ for $n=1$.

Therefore we have
\begin{equation}
\int_0^1 dx g_1^\gamma(x,Q^2,P^2)=
-\frac{3\alpha}{\pi}\sum_{i=1}^{n_f}{e_i}^4
+{\cal O}(\alpha_s). \label{virtual1}
\end{equation}
We can go a step further to ${\cal O}(\alpha_s)$ QCD corrections
which turn out to be \cite{SU99}
\begin{eqnarray}
&& \int_0^1dx g_1^\gamma(x,Q^2,P^2)\hspace{3cm}\nonumber\\
&&=-\frac{3\alpha}{\pi}
\left[\sum_{i=1}^{n_f}e_i^4\left(1-\frac{\alpha_s(Q^2)}{\pi}\right)
\right.\hspace{1cm}\nonumber\\
&&\hspace{1.0cm}-\left.\frac{2}{\beta_0}(\sum_{i=1}^{n_f}e_i^2)^2\left(
\frac{\alpha_s(P^2)}{\pi}-\frac{\alpha_s(Q^2)}{\pi}\right)\right]\nonumber\\
&&+{\cal O}(\alpha_s^2).\hspace{3cm} \label{virtual2}
\end{eqnarray}
This result coincides with the one obtained by Narison, Shore and 
Veneziano in ref.\cite{NSV93}, apart from the overall sign for the definition
of $g_1^\gamma$.

It would be extremely interesting to explain the transition from the vanishing
sum rule (\ref{real}) for the real photon to the non-vanishing result
(\ref{virtual1}) or (\ref{virtual2}) for the virtual photon .

\section{POSITIVITY CONSTRAINTS}

Application of the Cauchy-Schwarz inequality \cite{S95,ST98} to
the above photon helicity amplitudes leads to
the following positivity bound \cite{SSU01}:
\begin{equation}
|W(a,b|a',b')|\leq \sqrt{W(a,b|a,b)W(a',b'|a',b')}.
\end{equation} 
More explicitly, we obtain the following three
positivity constraints:
\begin{eqnarray}
&&\hspace{-0.7cm}|W(1,1|-1,-1)|\leq W(1,1|1,1)\\
&&\hspace{-0.7cm}|W(1,1|0,0)|\leq \sqrt{W(1,1|1,1)W(0,0|0,0)}\\
&&\hspace{-0.7cm}|W(1,0|0,-1)|\leq \sqrt{W(1,0|1,0)W(0,1|0,1)}.
\end{eqnarray}
In terms of the structure functions (\ref{structurefn}),
the positivity constraints read
\begin{eqnarray}
&&\hspace{-0.7cm}\Bigl|W_{ TT}^\tau \Bigr|\leq
\left(W_{ TT}+W_{ TT}^a\right)\label{realpositiv}\\
&&\hspace{-0.7cm}\Bigl|W_{ TS}^\tau +W_{ TS}^{\tau a}  \Bigr|\leq
\sqrt{(W_{ TT}+W_{ TT}^a)W_{ SS}}\\
&&\hspace{-0.7cm}\Bigl| W_{ TS}^\tau -W_{ TS}^{\tau a} \Bigr|\leq
\sqrt{W_{ TS}W_{ ST}}.
\end{eqnarray}

For the real photon target, $P^2=0$, the number of independent
structure functions or helicity amplitudes reduces to four, which
are \cite{SSU01}
\begin{eqnarray}
&&\hspace{-0.7cm} W_{TT}=\frac{1}{2}F_1^\gamma,\quad
 W_{ST}=\frac{1}{2x}F_L^\gamma, \quad
 W_{TT}^\tau=2W_3^\gamma,\nonumber\\
&& \hspace{-0.7cm}W_{TT}^a=\frac{1}{2}g_1^\gamma
\end{eqnarray}
and we have only one positivity constraint (\ref{realpositiv}).

The virtual photon structure functions $F_2^\gamma$, 
$F_L^\gamma$, and $g_1^\gamma$
are related to $W_i$'s as
\begin{eqnarray}
&&\hspace{-0.7cm}F_1^\gamma(x,Q^2,P^2)=2\left[ W_{TT}-\frac{1}{2}W_{TS}\right] \nonumber \\
&&\hspace{-0.7cm}F_2^\gamma(x,Q^2,P^2)=\frac{2x}{{\widetilde\beta}^2} \left[
   W_{TT}+W_{ST}-\frac{1}{2}W_{SS}\right.\nonumber\\
&&\left.\hspace{4.8cm} -\frac{1}{2}W_{TS}  \right] \nonumber \\
&&\hspace{-0.7cm}F_L^\gamma(x,Q^2,P^2)={\widetilde\beta}^2F_2^\gamma-xF_1^\gamma,  \nonumber\\
&&\hspace{-0.7cm}g_1^\gamma(x,Q^2,P^2)=\frac{2}{{\widetilde\beta}^2}\left[W^a_{TT}-
\frac{(P^2Q^2)^{1/2}}{w}W^{\tau a}_{TS}
     \right]  \nonumber\\
&&\hspace{-0.7cm}g_2^\gamma(x,Q^2,P^2)=-\frac{2}{{\widetilde\beta}^2}\left[W^a_{TT}-
\frac{w}{(P^2Q^2)^{1/2}}W^{\tau a}_{TS}
     \right]  \nonumber\\
&&\hspace{-0.7cm}\mbox{with}\quad {\widetilde\beta}=\left(1-\frac{P^2Q^2}{w^2}\right)^{1/2} \mbox{and} \ w=p\cdot q.
\end{eqnarray}
For $\widetilde\beta  \approx 1$
\begin{eqnarray}
&&\hspace{-0.7cm}W_{TT}(x,Q^2,P^2)\approx\frac{1}{2} F_1^\gamma(x,Q^2,P^2)\nonumber\\
&&\hspace{0cm}=
\frac{1}{2x}\left\{F_2^\gamma(x,Q^2,P^2)-F_L^\gamma(x,Q^2,P^2)\right\}
\nonumber \\
&&\hspace{-0.7cm}W^a_{TT}(x,Q^2,P^2)\approx \frac{1}{2}g_1^\gamma(x,Q^2,P^2).
\end{eqnarray}

Now let us see if the first positivity constriant (\ref{realpositiv}) is
satisfied in the pQCD results.
\begin{eqnarray}
&&\hspace{-0.7cm}\Bigl|W_{ TT}^\tau(x,Q^2,P^2) \Bigr|\nonumber\\
&&\mathop{<}_\sim
\frac{1}{2}\left[F_1^\gamma(x,Q^2,P^2)+g_1^\gamma(x,Q^2,P^2)\right].
\label{ineq}
\end{eqnarray}

\vspace{-1cm}
\begin{figure}[htbp]
\vspace{-0.3cm}
\begin{center}
\epsfxsize=7cm
\hspace*{0cm}
\ \epsfbox{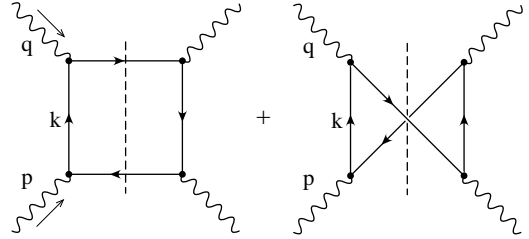}
\vspace{-1cm}
\caption{The Box diagram contributions}
\end{center}
\end{figure}
\vspace{-1cm}

The parton model Box diagram (Fig.4) calculation 
for $W_{TT}^\tau(x,Q^2,P^2)=2W_3^\gamma$ gives
the LO QCD result up to ${\cal O}(\alpha_s)$ as
\begin{eqnarray}
&&\hspace{-0.7cm}W_{TT}^\tau(x,Q^2,P^2)\nonumber\\
&&=\frac{\alpha}{2\pi}\delta_{\gamma}
\left[(-2x^2)+{\cal O}(\alpha_s(Q^2)) \right].
\end{eqnarray}

\begin{figure}[htbp]
\vspace{-3cm}
\vspace{-0.3cm}
\begin{center}
\epsfxsize=8cm
\hspace*{-0.7cm}
\ \epsfbox{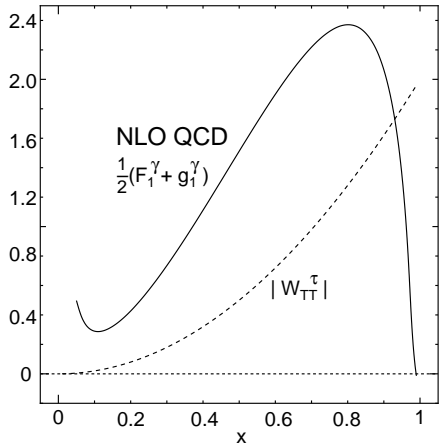}
\vspace{-3cm}
\caption{NLO QCD result of $\frac{1}{2}[F_1^\gamma
+g_1^\gamma] $ vs. LO result of $W_{TT}^\tau$ in
units of ($\alpha/2\pi)\delta_\gamma$, for $Q^2=30$GeV$^2$,
$P^2=1$GeV$^2$ with $\Lambda=0.2$GeV and $n_f=3$.}
\end{center}
\end{figure}
In Fig. 5 we have plotted the NLO QCD result of
$\frac{1}{2}[F_1^\gamma+g_1^\gamma]$ and the LO 
result of $|W_{TT}^\tau|$ as functions of $x$ for
the case $P^2/Q^2=1/30$ with the number of active
flavors, $n_f=3$.

We observe that the inequality (\ref{ineq}) is
satisfied for almost all the allowed $x$ region
except near $x_{\rm max}=1/(1+P^2/Q^2)$ 
($\approx 0.968$ for $P^2/Q^2=1/30$).
The reason of why the inequality is not
satisfied near $x_{\rm max}$ is as follows.
We notice that the graph of $\frac{1}{2}(
F_1^\gamma+g_1^\gamma)$ falls rapidly as
$x \rightarrow x_{\rm max}$. In the QCD
parton picture, this is due to the total
momentum conservation of the partons inside
the photon. From the large $n$ behavor of
the structure functions $F_1^\gamma$ and
$g_1^\gamma$, they vanish like $-1/\ln(1-x)$
as $x \rightarrow 1$. The NLO QCD effects
further suppress $F_1^\gamma$ and $g_1^\gamma$ 
at large $x$. While, the LO QCD result of $W_{TT}^\tau$
is just same as the massless quark parton model,
where the power corrections of $P^2/Q^2$ is
neglected. Hence $|W_{TT}^\tau|$ in Fig. 5 increases
as a function of $x$ like $x^2$ and violates the 
inequality in the vicinity of $x_{\rm max}$.
The {\it physical} $W_{TT}^\tau$ should vanish as
$x \rightarrow x_{\rm max}$, which is realized
in the case of parton model with massive quark.
This implies the necessity of introducing the 
quark mass effects for the photon coefficient 
function.

\section{CONCLUSION}

To summarize we have investigated the aspects of the structure 
functions of the virtual photon for the most general case.
The virtual photon structure 
could be studied in future $ep$ and $e^+e^-$ colliders and
provides a good testing ground for the structure of the 
current product.

We have discussed the sum rule for the $g_1^\gamma$ for
the virtual photon target. It would be an intriguing
question how we can understand the transition from vanishing 
1st moment for real photon to non-vanishing one 
for virtual photon.

For the eight independent structure functions of the 
virtual photon we have derived three positivity constraints 
and discussed if the first inequality is satisfied in the QCD
parton model computation of $F_1^\gamma$, $g_1^\gamma$ and 
$W_{TT}^\tau$. 

For another polarized structure function $g_2^\gamma$, the
twist-3 effects are important, where we have to
solve the mixing problem for the full QCD analysis \cite{KS02}.
There remain a number of virtual photon structure functions
yet to be studied.

\vspace{0.5cm}
\leftline{\bf ACKNOWLEDGMENTS}
\vspace{0.5cm}
The author thanks the Organizers of RADCOR/Loops and Legs
2002 for the kind invitation, and
Ken Sasaki and Jacques Soffer for the collaboration
on the topics presented here.

\end{document}